\documentclass[prl,twocolumn,amsmath,amssymb,showpacs]{revtex4}
\usepackage{times}
\usepackage{graphicx}
\usepackage{dcolumn}
\usepackage{bm}
\def \bea{\begin{eqnarray}}
\def \eea{\end{eqnarray}}

\begin{document}


\title{Scaling laws for the response of nonlinear elastic media with implications for cell mechanics}
\author{Yair Shokef$^{1,2}$ and Samuel A. Safran$^2$}
\affiliation{
$^1$School of Mechanical Engineering, Tel-Aviv University, Tel-Aviv 69978, Israel\\
$^2$Department of Materials and Interfaces, Weizmann Institute of Science, Rehovot 76100, Israel}

\begin{abstract}

We show how strain stiffening affects the elastic response to internal forces, caused either by material defects and inhomogeneities or by active forces that molecular motors generate in living cells. For a spherical force dipole in a material with a strongly nonlinear strain energy density, strains change sign with distance, indicating that even around a contractile inclusion or molecular motor there is radial compression; it is only at long distance that one recovers the linear response in which the medium is radially stretched. Scaling laws with irrational exponents relate the far-field renormalized strain to the near-field strain applied by the inclusion or active force. 

\end{abstract}



\maketitle


The response of elastic media to internal forces is an important factor that governs the physics of defects and inclusions in solids and composite materials~\cite{Eshelby}. Theory has shown that internal forces may be modeled as force dipoles that generate strains and stresses that can lead to interesting collective effects~\cite{Wagner_Horner_Safran_graphite}. Recently, similar ideas were applied to the mechanical response of biological cells~\cite{Saif} where the resulting physical phenomena~\cite{cell_mech_3,Zemel_Rehfeldt} are also associated with cell function~\cite{engler}.

The cytoskeleton of living cells contains molecular motors that consume ATP and produce nonequilibrium forces by which cells attach to and pull on their surroundings. This mechanical interaction relates to many aspects of cellular function, from cell spreading and proliferation to stem-cell differentiation and tissue development~\cite{engler}. Force is generated in cells by myosin motors that pull on the cross-linked actin filaments comprising the cytoskeleton. Most theoretical studies model these forces via the activity of force-dipoles that are embededded in a material described by linear elasticity~\cite{DeSafranPRE2008}. These forces can act within the cytoskeleton resulting in ordering of its actin filaments~\cite{Zemel_Rehfeldt,Friedrich_Buxboim}. On a larger scale, the entire, contractile cell~\cite{Balaban} can be represented as a force dipole that deforms its extracellular environment to produce strains and stresses that result in effective elastic interactions with other cells~\cite{cell_mech_5}. However, both the cytoskeleton and the extracellular matrix comprise cross-linked, semi-flexible polymeric filaments that respond linearly (with elastic constants that are stress independent) only for small stresses. For stresses beyond a critical value, these gels show a nonlinear response with power-law stiffening of the elastic moduli with increasing stress~\cite{strain_stiffening}. Indeed, nonlinear behavior was measured in the elastic and viscous response of biological cells~\cite{Ott}.


\begin{figure}
\includegraphics[width=0.99\columnwidth]{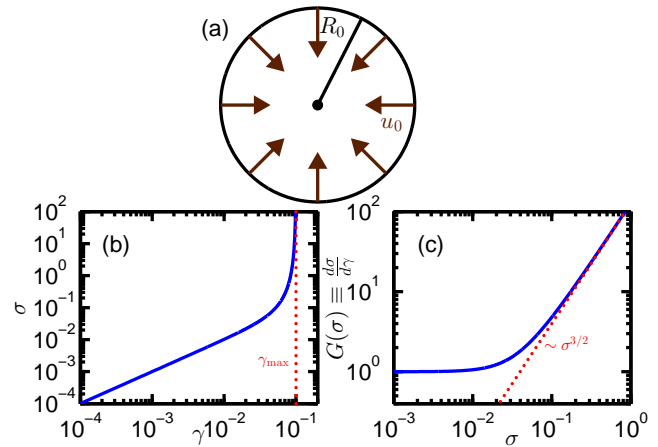}
\caption{a) Spherical force dipole generates radial displacement $u_0$ at radius $R_0$. b-c) Constitutive relations for simple shear: Eq.~(\ref{eq:knowles}) with $n=-1$, $b=100$, $\mu=1$: b) Shear stress vs strain diverges at $\gamma_{\rm max}$. c) Differential modulus vs shear stress asymptotes to $G \propto \sigma^\beta$.}
\label{fig:setup}
\end{figure}

We report on the first step in theoretical understanding of how the nonlinear mechanical properties of an elastic medium can radically change the strains and stresses generated by internal forces in a long-range manner. Our results are important for the understanding of deformations induced by inclusions in nonlinear elastic media, active forces in in-vitro acto-myosin gels;  they also have implications for the interactions of acto-myosin segments in cells as well as interactions of cells in nonlinear biopolymer media~\cite{Winer}. We consider an infinite, homogeneous, and isotropic compressible elastic material whose response is linear for small stress, but shows strain stiffening, with a power-law increase of its differential shear modulus with increasing stress, see Fig.~\ref{fig:setup}c. We analyze the response of this medium to a spherical force dipole which exerts an isotropic radial force on its surface, see Fig.~\ref{fig:setup}a. This can represent a spherical inclusion in a gel and is also motivated by the forces exerted by relatively symmetric~\cite{pompe} contractile, non-motile cells such as endothelial cells for which adhesion sites are uniformly distributed on their perimeter~\cite{Dikla_adv_mater}. Our model predicts analytic scaling laws as well as numerical solutions for the strain that the force dipole generates in the medium, as well as its associated energy cost. Our results suggest that the long-range interactions of such force dipoles is significantly modified by the nonlinearity of the medium~\cite{cavitation}.

For simplicity, we first describe the force dipole by the radial displacement $u_0$ that it generates at a distance $R_0$ from its center. We will later invert the problem to obtain the force applied at $R=R_0$. The deformation decays with the distance $R$ from the center of the dipole. At $R \gg R_0$ the stress is small enough so that the distance dependence of the displacement is identical to a linear medium, $u(R) \approx u_{\rm eff} (R_0/R)^2$, but with a coefficient $u_{\rm eff}$ that can differ significantly from that of a linear elastic medium for which  $u_{\rm eff} \equiv u_0$. We find that for weak nonlinearity (small $u_0$), the deviation from the linear solution scales as $u_{\rm eff}-u_0 \propto u_0^3$, whereas for strong nonlinearity (large $u_0$) $u_{\rm eff}$ scales as $u_{\rm eff} \propto u_0^\theta$, with the irrational exponent $\theta = \frac{6}{5-\sqrt{15}} \approx 5.32$. These scaling laws induce related, nontrivial scaling relations between the force dipole moment, the induced strain and stress, and the total elastic energy required to generate the deformation, which are the focus of recent traction force microscopy experiments~\cite{pompe}.


A gel that stiffens as shear stress is increased may be modeled by the following elastic energy density functional~\cite{knowles}:
\bea
W = \frac{\mu}{2b}\left\{\left[1+\frac{b}{n}\left(\bar{I}_1-3\right)\right]^n-1\right\} + \frac{K}{2} \left( J - 1 \right)^2. \label{eq:knowles}
\eea
This applies to a system in which the strain is not necessarily small so that $J = \det({\bf F})$ measures the compression with $F_{ij}=\frac{\partial x_i}{\partial X_j}$ the deformation gradient tensor, $\vec{X}$ the reference position and $\vec{x}$ the deformed position. The shear deformation is given by $\bar{I}_1 \equiv I_1 / J^{2/3}$ with $I_1 = {\rm tr}({\bf B})$ and $B_{ij}=F_{ik}F_{jk}$ the left Cauchy-Green strain tensor~\cite{Bower}. For vanishingly small values of the dimensionless parameter $b$ characterizing the nonlinearity, (\ref{eq:knowles}) yields a compressible neo-Hookean form $W = \frac{\mu}{2}\left(\bar{I}_1-3\right) + \frac{K}{2} \left( J - 1 \right)^2$, and the Cauchy stress tensor $\sigma_{ij} = \frac{1}{J} F_{ik}\frac{\partial W}{\partial F_{kj}}$ obtains the form $\boldsymbol{\sigma} = \frac{\mu}{J^{5/3}}\left({\bf B} - \frac{I_1}{3}{\bf 1}\right) + K(J-1) {\bf 1}$, with ${\bf 1}$ the unit tensor. For small deformations, $B_{ij} \approx \delta_{ij} + 2\epsilon_{ij}$, with $\epsilon_{ij} = \frac{1}{2}\left(\frac{\partial u_i}{\partial X_j} + \frac{\partial u_j}{\partial X_i}\right)$ the linear strain tensor, and $\vec{u} = \vec{x}-\vec{X}$ the displacement. In this limit, $I_1 \approx 3 + 2 {\rm tr}(\boldsymbol{\epsilon})$ and $J \approx 1 + {\rm tr}(\boldsymbol{\epsilon})$, thus the constitutive relations reduce to Hooke's law, $\boldsymbol{\sigma} = 2 \mu \boldsymbol{\epsilon} + \left( K - \frac{2}{3}\mu \right){\rm tr}(\boldsymbol{\epsilon}) {\bf 1}$, with shear modulus $\mu$ and bulk modulus $K$. 

For simple shear ($x=X+\gamma Z$, $y=Y$, $z=Z$), (\ref{eq:knowles}) gives $\sigma_{xz} = \mu \gamma \left(1+\frac{b}{n}\gamma^2\right)^{n-1}$. In the limit of small shear, the differential shear modulus $G \equiv \frac{d\sigma_{xz}}{d\gamma}$ is constant and equal to $\mu$. For $n<0$ and $b>0$, $\sigma_{xz}$ diverges as the shear increases and the strain $\gamma$ approaches $\gamma_{\rm max}=\sqrt{-\frac{n}{b}}$, while the shear modulus stiffens as  $G \propto \sigma^\beta$ with $\beta=\frac{n-2}{n-1}$, see Fig.~\ref{fig:setup}b-c. 

Thus $b$ determines the maximal strain $\gamma_{\rm max}$, which is experimentally known to be related to the concentration of actin, collagen or crosslinkers in biological gels~\cite{strain_stiffening}. Preliminary estimates based on the experimental data yield values $2<b<400$. Theoretically, strong nonlinearity may be introduced at extremely small strain by taking arbitrarily large values of $b$, which induce arbitrarily small values of $\gamma_{\rm max}$~\cite{rajagopal}. Since many biopolymer gels are nearly incompressible, we assume the compression is small and treat it linearly. $n$ determines the exponent $\beta$ that quantifies the strain-stiffening behavior of $G$ vs $\sigma$. The nonlinearity of semi flexible chains implies that $\beta=\frac{3}{2}$~\cite{strain_stiffening}, which is obtained by taking $n=-1$. 


For the spherically symmetric geometry of Fig.~\ref{fig:setup}a, the deformation is described by the radial displacement $u(R)$. We focus on extremely nonlinear materials with $b \gg 1$, for which the nonlinearity enters at extremely small strain. Here, the strain scales as $u/R$, therefore $u/R \ll 1$ and it is tempting to keep only the terms linear in $u$. However, since $b u^2$ is not necessarily small, we keep both the linear terms and terms of order $bu^3$. We derive the stress tensor from the derivative of (\ref{eq:knowles}) with respect to the deformation, employ this small-strain approximation, and eventually obtain the following equation of mechanical equilibrium in which the divergence of the stress vanishes for $R>R_0$ where there are no internal forces~\cite{Bower}:
\bea
\frac{3K}{4\mu} L +\left[1+\frac{A}{n}\left(\tilde{I}_1-3\right)\right]^{n-2}\left[ L + A \left(C_1+\frac{C_2}{n}\right)\right] =0 \label{eq:the_equation}
\eea
with
\bea
\tilde{I}_1-3 &=& -\frac{4}{9}\left[2\left(\frac{d\tilde{u}}{d\tilde{R}}\right)^2 +6\frac{\tilde{u}}{\tilde{R}}\cdot\frac{d\tilde{u}}{d\tilde{R}} -3\frac{\tilde{u}^2}{\tilde{R}^2}\right] \label{eq:I1m3} ,
\eea
$L \equiv \frac{d^2\tilde{u}}{d\tilde{R}^2}+\frac{2}{\tilde{R}}\frac{d\tilde{u}}{d\tilde{R}}-\frac{2\tilde{u}}{\tilde{R}^2}$,
$C_1 \equiv \frac{d\tilde{I}_1}{d\tilde{R}}\left(\frac{d\tilde{u}}{d\tilde{R}}-\frac{\tilde{u}}{\tilde{R}}\right)$,
$C_2 \equiv \left(\tilde{I}_1-3\right) L - C_1$, $\tilde{u} \equiv u/u_0$, $\tilde{R} \equiv R/R_0$, and $A \equiv b u_0^2 / R_0^2$.

\begin{figure}
\includegraphics[width=0.94\columnwidth]{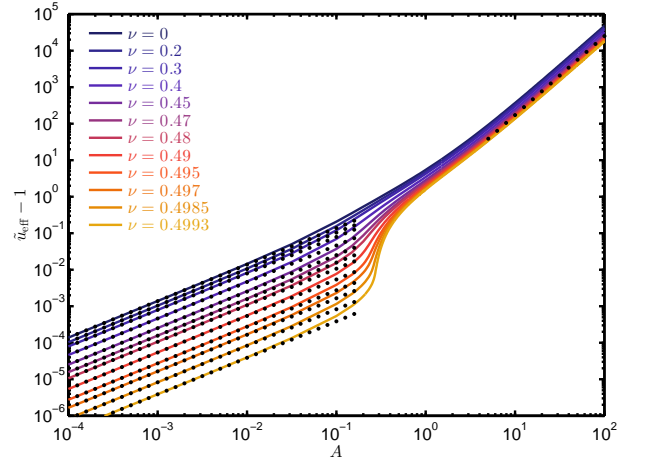}
\caption{(Color online) Normalized effective far-field displacement vs dimensionless parameter $A \equiv b u_0^2 / R_0^2$ quantifying the nonlinearity in the system. Color lines indicate numerical solutions of Eq.~(\ref{eq:the_equation}) for different Poisson ratios, and $n=-1$. Dotted lines are theoretical solutions: Eq.~(\ref{eq:u_eff_small_A}) at small $A$ and Eq.~(\ref{eq:ueff_large_A}) at large $A$.}
\label{fig:ueff}
\end{figure}

\emph{Weak nonlinearity, $A \ll 1$:} The linear ($A \rightarrow 0$) solution is $\tilde{u}(\tilde{R})= 1/\tilde{R}^2$. For $A\tilde{u}^2 \ll 1$, linearization leads to 
\bea
\tilde{u} = \frac{\tilde{u}_{\rm eff}}{\tilde{R}^2}\left(1+\frac{D}{\tilde{R}^6}\right), \label{eq:small_A}
\eea
with $D \equiv -\frac{56}{81} \frac{1-2\nu}{1-\nu} \left(1-\frac{1}{n}\right) A \tilde{u}_{\rm eff}^2$, and $\nu\equiv\frac{3K-2\mu}{2(3K+\mu)}$ the Poisson ratio. For any $A$ this is valid for large enough $\tilde{R}$. For $A \ll 1$ it is valid already from $\tilde{R}=1$, where $\tilde{u}(1)=1$. Thus 
\bea
\tilde{u}_{\rm eff} = 1 + \frac{56}{81} \frac{1-2\nu}{1-\nu} \left(1-\frac{1}{n}\right) A , \label{eq:u_eff_small_A}
\eea
implying that $u_{\rm eff}-u_0 \propto u_0^3$. Agreement with the numerical solution of Eq.~(\ref{eq:the_equation}) may be seen in Fig.~\ref{fig:ueff} to hold up to $A \approx 0.1$. 

\begin{figure}
\includegraphics[width=0.92\columnwidth]{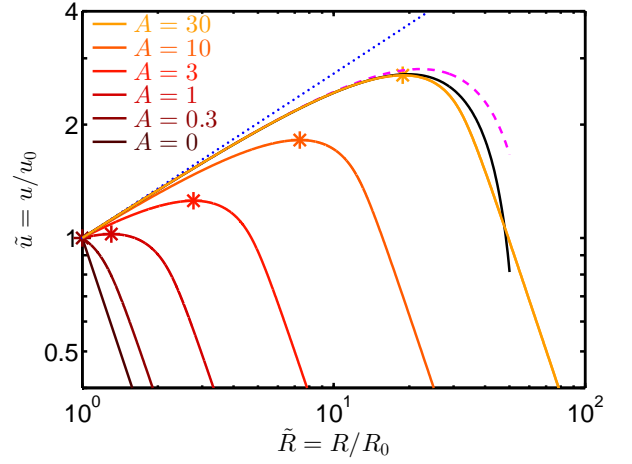}
\caption{(Color online) Normalized displacement vs normalized radius for $n=-1$, Poisson ratio $\nu=0.4$, and various values of the dimensionless parameter $A \equiv bu_0^2/R_0^2$ characterizing the nonlinearity in the system. Colored solid lines are results of numerically solving the equation of mechanical equilibrium (\ref{eq:the_equation}). Stars denote the locations of the maxima. Dotted blue line is Eq.~(\ref{eq:u_approx_1}), dashed magenta line is Eq.~(\ref{eq:u_approx_2}) and solid black line is higher order correction~\cite{SI}.}
\label{fig:u}
\end{figure}

\emph{Strong nonlinearity, $A \gg 1$:} The deformation has a shear component, $\bar{I}_1-3$, and a compressive component, $J$. In our model, only the resistance to shear stiffens with increasing stress. Hence, for strong nonlinearity, the deformation energy will be dominated by shear, so that energy minimization implies that $\tilde{I}_1-3$ should be minimal. A zero shear deformation is derived from (\ref{eq:I1m3}) with
\bea
\tilde{u} = \tilde{R}^\alpha, \label{eq:u_approx_1}
\eea
and $\alpha = \frac{-3 + \sqrt{15}}{2} \approx 0.44$. The reason for this irrational exponent (and for subsequent irrational scaling exponents) arises from the homogeneous nature of (\ref{eq:I1m3}) that results in a quadratic equation for $\alpha$. However, (\ref{eq:u_approx_1}) is not an exact solution of Eq.~(\ref{eq:the_equation}) for all values of $\tilde{R}$, although Fig.~\ref{fig:u} shows that indeed as $A$ increases, the deformation approaches (\ref{eq:u_approx_1}) when $\tilde{R}$ is close to 1 and the displacement is large. Eq.~(\ref{eq:the_equation}) can be satisfied to higher order in $\tilde{R}$ by adding a correction to $\tilde{u}(\tilde{R})$ of order $\frac{1}{A}$ that causes $1+\frac{A}{n}\left(\tilde{I}_1-3\right)$ to vanish to leading order:
\bea
\tilde{u} = \tilde{R}^{\alpha} + \frac{9n}{80 \alpha A}\left( \tilde{R}^{2-\alpha} - \tilde{R}^{\alpha} \right). \label{eq:u_approx_2}
\eea
This provides a good approximation for $\tilde{u}(\tilde{R})$ and in particular reproduces the nonmonotonic behavior in the numerical solutions, see Fig.~\ref{fig:u}. In a linear medium, displacements decay monotonically. Namely, for a contractile force ($u_0<0$), the radial strain satisfies $\frac{du}{dR}<0$ for all $R>R_0$, and $\frac{du}{dR}>0$ only for $R<R_0$. In a nonlinear medium, $\frac{du}{dR}>0$ in a large region near the inclusion ($R<R_*$) and $\frac{du}{dR}<0$ only for large values of $R>R_*$ where Eq. (\ref{eq:small_A}) is valid. One way to understand this is that for nonlinear elasticity there is a strong penalty on shear strains above a critical level, and extending the region in which $\frac{du}{dR}>0$ helps to reduce the shear strain. Finally, $R_*$ represents an effective size of the force dipole, whose strain field extends over a distance much larger than its size $R_0$.

Defining $\tilde{R}_*$ as the position of the maximum of (\ref{eq:u_approx_2}),
\bea
\tilde{R}_* \equiv \frac{R_*}{R_0} = \left[ - \frac{80\alpha^2 A}{9n(2-\alpha)} \right]^{\frac{1}{2-2\alpha}} \label{eq:Rstar} ,
\eea
we find that for $A \gg 1$, higher order corrections to $\tilde{u}(\tilde{R})$ are functions of the scaled distance $\tilde{R}/\tilde{R}_*$~\cite{SI}. The increasingly good agreement of these corrections with the numerical solutions as the nonlinearity $A$ increases is seen in Fig.~\ref{fig:u}. For large $A$, $\tilde{R}_* \gg 1$ or $R_* \gg R_0$. The increase in $R_*$ with $A$ suggests that even at relatively large distances from $R_0$, the inclusion cannot be treated as a point dipole.  This may have important consequences for the interactions between force dipoles in a nonlinear medium. Our results suggest that these interactions will be significantly and qualitatively increased for large values of $A$. Such interactions at surprisingly long distances are consistent with recent experimental measurements of biological cells on nonlinear elastic substrates~\cite{Winer}.

At long distances the displacements obey $\tilde{u}(\tilde{R}) = \tilde{u}_{\rm eff}/\tilde{R}^2$. The crudest way to obtain $\tilde{u}_{\rm eff}$ is by matching this asymptotic large-$R$ form with (\ref{eq:u_approx_1}) derived for small $R$. Thus
\bea
\tilde{u}_{\rm eff} = \tilde{R}_*^{2+\alpha} \propto A^{\frac{2+\alpha}{2-2\alpha}} \approx A^{2.17} . \label{eq:ueff_large_A}
\eea
As seen in Fig.~\ref{fig:ueff}, the scaling with $A$ is in excellent agreement with the numerics while the prefactor differs slightly. Higher order terms both in the small-$\tilde{R}$ solution (\ref{eq:u_approx_2}) or the higher-order corrections given in~\cite{SI}, and in the large-$\tilde{R}$ solution (\ref{eq:small_A}) yield a more precise prediction for $u_{\rm eff}$. However, the scaling of $\tilde{u}_{\rm eff}$ with $\tilde{R}_*$ (or with $A$) remains unaltered, and even with such more complicated matching schemes there is still some discrepancy in the prefactors compared to the numerical solution. Noting that $\tilde{u}_{\rm eff} = u_{\rm eff} / u_0$, and $A \propto u_0^2$, we rewrite (\ref{eq:ueff_large_A}) as $u_{\rm eff} \propto u_0^\theta$ with $\theta = \frac{3}{1-\alpha} = \frac{6}{5-\sqrt{15}} \approx 5.32$. It will be interesting to test this relation experimentally by inferring the near- and far-field displacements proportional to $u_0$ and $u_{\rm eff}$ from the motion of fluorescent beads in nonlinear media deformed by contractile cells.


\emph{Stress:} So far we considered the boundary-value problem and solved the displacement field $u(R)$ given a certain displacement $u_0$ at radius $R_0$. The physically and biologically interesting questions have to do with the relations between the total force $4 \pi R_0^2 \sigma_0$ applied at $R_0$ and the deformations it generates, where $\sigma_0 \equiv \sigma_{rr}(R_0)$. In this context, two separate questions should be asked. One with respect to the deformation $u_0$ at $R_0$ and the second with respect to the deformation $u_{\rm eff}$ felt at $R \gg R_0$. 

In the linear case ($A=0$), $\sigma_0^{\rm lin} = - \frac{4\mu u_0}{R_0} \propto u_0 \equiv u_{\rm eff}$. For weak nonlinearity ($A \ll 1$), we find~\cite{SI} that the correction to the stress compared with the linear solution is linear in A:
\bea
\sigma_0 - \sigma_0^{\rm lin} \propto \frac{\mu u_0}{R_0} A \propto u_0^3 \propto u_{\rm eff} - u_0 . \label{eq:sigma_small_A}
\eea
For strong nonlinearity ($A \gg 1$), we find~\cite{SI} that the stress increases with $A$ with an irrational exponent related to $\alpha$:
\bea
\sigma_0 \propto \frac{\mu u_0}{R_0} A^{\frac{2+\alpha}{2-2\alpha}} \propto u_0^\theta \propto u_{\rm eff} . \label{eq:sigma_large_A}
\eea
In (\ref{eq:sigma_small_A}) and (\ref{eq:sigma_large_A}) we used the previously established connections between $u_{\rm eff}$ and $u_0$ in these two limits. To summarize, we find nontrivial scaling relations between the applied stress $\sigma_0$ and the displacement $u_0$ generated where the force is applied ($R=R_0$), as well as nontrivial scaling relations between $u_0$ and the effective displacement $u_{\rm eff}$ felt far away from the force dipole. However, taken together, these relations reduce to a linear relation between $\sigma_0$ and $u_{\rm eff}$. Thus the renormalization of $u_{\rm eff}$ of the linear solution for $R \gg R_*$ also characterizes the renormalization of the stress applied at $R=R_0$. Interestingly, the same relation $\sigma_0 \propto u_{\rm eff}$ is true for both weak ($A \ll 1$) and for strong ($A \gg 1$) nonlinearities. Such renormalization is similar to that investigated for charged colloids~\cite{Alexander}

These results may help resolve the question of whether cells sense stress or strain~\cite{stress_or_strain} by measurements on identical cells placed on nonlinear gels with different values of $\mu$ (and $b$). Since $\sigma_0 \propto \mu u_{\rm eff}$, if the cells exert constant stress, one would measure $u_{\rm eff} \propto \frac{1}{\mu}$. If, on the other hand, cells exert constant strain, Eq.~(\ref{eq:ueff_large_A}) leads to $u_{\rm eff} \propto b^{\frac{2+\alpha}{2-2\alpha}}$. Experimentally, $\mu$ and $b$ are not independent, but are roughly related by $\mu \propto \gamma_{\rm max}^{-a} \propto b^{\frac{a}{2}}$, with $3 <a< 6$~\cite{strain_stiffening}. This leads to $u_{\rm eff} \propto \mu^\eta$ with $0.7 < \eta=\frac{2+\alpha}{(1-\alpha)a} < 1.4$, which is easily distinguishable from $u_{\rm eff} \propto \frac{1}{\mu}$.


We now connect our results to additional experimentally measurable quantities~\cite{pompe}, and briefly discuss how the total energy stored in the elastic deformation (the self energy of the force dipole), scales with the force and displacement. In our formalism this energy is given by $U = \int W d^3\vec{r}$. However this is also equal to the work performed by the active forces starting from an undeformed state. Since the external forces are applied only at $R_0$, this simplifies to $U = 2\pi R_0^2 \sigma_0 u_0 \propto \sigma_0 u_0$. In the linear case, $U \propto \sigma_0^2$ (or $u_0^2$). For strong nonlinearity ($A \gg 1$), we obtain $U \propto \sigma_0^{1+1/\theta}$ (or $u_0^{\theta+1}$) $\approx \sigma_0^{1.19}$ (or $u_0^{6.32}$). A direct comparison with the nonlinear scaling laws deduced from traction-force microscopy experiments~\cite{pompe} is beyond the scope of this paper, since these measured displacements assumed linear elasticity for calculating force and self energy. Moreover, in the above analysis we assumed a fixed spatial size $R_0$ for the force dipoles and studied the dependence only on the dipole strength as parameterized by $u_0$ or $\sigma_0$, whereas in a more realistic model there may be correlations between the radius of the spherical dipole (or cell size) and its activity. Nonetheless, our predicted irrational scaling exponents should show up in such measurements.


In summary, we predict scaling laws for the response of a nonlinear elastic medium to an internal, spherical force dipole.  We find that the nonlinearity changes the induced strain from one that is constant in sign to a strain that changes sign at a distance $R_*$ that for strong nonlinearities can be very large compared to the size of the dipole. We also predict scaling laws for the displacement that behaves as a power law with irrational exponent. Moreover, we identify nontrivial scaling relations between the applied force ($\sigma_0$) and the deformations generated at the dipole surface ($u_0$), and between these deformations and the deformations felt at long distances ($u_{\rm eff}$). However, surprisingly, we find that the stress applied by the spherical dipole can always be written as a linear function of the renormalized displacement $u_{\rm eff}$ characterizing the strains induced at very large distances. The magnitude of the effects we find can significantly enhance the interactions between force dipoles. This will have important implications for cell-cell interactions in nonlinear elastic media as already suggested by the experiments in~\cite{Winer}.


We thank Eran Bouchbinder, Benjamin Friedrich, Doron Kushnir, Itamar Procaccia and Ulrich Schwarz for helpful discussions. We thank the ISF and the Perlman Family Foundation for financial support.



\begin{thebibliography}{100}

\bibitem{Eshelby} J.D. Eshelby, Proc. R. Soc. London, Ser. A {\bf 241}, 376 (1957).

\bibitem{Wagner_Horner_Safran_graphite} 
H. Wagner and H. Horner, Adv. Phys. {\bf 23}, 587 (1974);
S.A. Safran and D.R. Hamann, Phys. Rev. Lett. {\bf 42}, 1410 (1979).

\bibitem{Saif} X. Tang, P. Bajaj, R. Bashir and T.A. Saif, Soft Matter {\bf 7}, 6151 (2011).

\bibitem{cell_mech_3}
U.S. Schwarz and S.A. Safran, Phys. Rev. Lett. {\bf 88} 048102 (2002);
R. De, A. Zemel and S.A. Safran, Nat. Phys. {\bf 3}, 655 (2007);
B.M. Friedrich and S.A. Safran, Europhys. Lett. {\bf 93}, 28007 (2011).

\bibitem{Zemel_Rehfeldt} A. Zemel, F. Rehfeldt, A.E.X. Brown, D.E. Discher, and S.A. Safran, Nat. Phys. {\bf 6}, 468 (2010).

\bibitem{engler} A.J. Engler, S. Sen, H.L. Sweeney, and D.E. Discher, Cell {\bf 126}, 677 (2006).

\bibitem{DeSafranPRE2008} R. De and S.A. Safran, Phys. Rev. E {\bf 78}, 031923 (2008).

\bibitem{Friedrich_Buxboim} B.M. Friedrich, A. Buxboim, D.E. Discher and S.A. Safran, Biophys J. {\bf 100}, 2706 (2011).

\bibitem{Balaban} N.Q. Balaban et al., Nat. Cell Biol. {\bf 3}, 466 (2001).

\bibitem{cell_mech_5} 
T. Korff and H.G. Augustin, J. Cell Sci. {\bf 112}, 3249 (1999);
I.B. Bischofs, S.A. Safran and U.S. Schwarz, Phys. Rev. E {\bf 69}, 021911 (2004);
I.B. Bischofs and U.S. Schwarz, Phys. Rev. Lett. {\bf 95}, 068102 (2005).

\bibitem{strain_stiffening} 
M.L. Gardel, J.H. Shin, F.C. MacKintosh, L. Mahadevan, P. Matsudaira, and D.A. Weitz, Science {\bf 304}, 1301 (2004);
C. Storm, J.J. Pastore, F.C. MacKintosh, T.C. Lubensky, and P.A. Janmey, Nature {\bf 435}, 191 (2005);
D. Vader, A. Kabla, D. Weitz, L. Mahadevan, PLoS ONE {\bf 4}, e5902 (2009).

\bibitem{Ott} P. Fern\'{a}ndez, P.A. Pullarkat, and A. Ott, Biophys. J. {\bf 90}, 3796 (2006).

\bibitem{Winer} J.P. Winer, S. Oake, and P.A. Janmey, PLoS ONE {\bf 4}, e6382 (2009).

\bibitem{pompe} T. Pompe, S. Glorius, T. Bischoff, I. Uhlmann, M. Kaufmann, S. Brenner, and C. Werner, Biophys. J. {\bf 97}, 2154 (2009).

\bibitem{Dikla_adv_mater} D. Raz-Ben Aroush and H.D. Wagner, Adv. Mater. {\bf 18}, 1537 (2006).

\bibitem{cavitation} For related work on cavitation in nonlinear media see: C.O. Horgan and D.A. Polignone, Appl. Mech. Rev. {\bf 48}, 471 (1995).

\bibitem{knowles} J.K. Knowles, Int. Jour. of Fracture {\bf 13}, 611 (1977).

\bibitem{Bower} A.F. Bower, Applied Mechanics of Solids (CRC Press) 2009. URL http://solidmechanics.org/

\bibitem{rajagopal}
K.R. Rajagopal, Math. Mech. Solids {\bf 16}, 122 (2010);
K.R. Rajagopal, Math. Comput. Appl. {\bf 15}, 506 (2010).

\bibitem{SI} See Supplemental Material.

\bibitem{Alexander} S. Alexander, P.M. Chaikin, P. Grant, G.J. Morales, P. Pincus, D. Hone, J. Chem. Phys. {\bf 80}, 5776 (1984).

\bibitem{stress_or_strain} R. De, A. Zemel, and S.A. Safran, Biophys. J. {\bf 94}, L29 (2008).

\end{thebibliography}
\end{document}